\documentclass[twocolumn,showpacs,preprintnumbers,amsmath,amssymb]{revtex4}
\usepackage{graphicx}% Include figure files
\usepackage{epsfig}
\usepackage{dcolumn}% Align table columns on decimal point
\usepackage{bm}% bold math

\def\be{\begin{equation}}
\def\ee{\end{equation}}
\def\ba{\begin{eqnarray}}
\def\ea{\end{eqnarray}}

\begin{document}

\title{Robertson-Schr\"odinger type formulation of Ozawa's noise-disturbance uncertainty principle}

\author{Catarina Bastos\footnote{E-mail: catarina.bastos@ist.utl.pt}}

\affiliation{Instituto de Plasmas e Fus\~ao Nuclear, Instituto Superior T\'ecnico Avenida Rovisco Pais 1, 1049-001 Lisboa, Portugal}

\author{Alex E. Bernardini\footnote{On leave of absence from Departamento de F\'isica, Universidade
Federal de S\~ao Carlos, PO Box 676, 13565-905, S\~ao Carlos, SP, Brasil. E-mail: alexeb@ufscar.br}, Orfeu Bertolami\footnote{Also at Instituto de Plasmas e Fus\~ao Nuclear, Instituto Superior T\'ecnico,
Avenida Rovisco Pais 1, 1049-001 Lisboa, Portugal. E-mail: orfeu.bertolami@fc.up.pt}}

\affiliation{Departamento de F\'isica e Astronomia, Faculdade de Ci\^encias da Universidade do Porto, Rua do Campo Alegre, 687,4169-007 Porto, Portugal}

\author{{Nuno Costa Dias and Jo\~ao Nuno Prata}\footnote{Also at Grupo de F\'{\i}sica Matem\'atica, UL,
Avenida Prof. Gama Pinto 2, 1649-003, Lisboa, Portugal. E-mail: ncdias@meo.pt, joao.prata@mail.telepac.pt}}
\affiliation{Departamento de Matem\'{a}tica, Universidade Lus\'ofona de
Humanidades e Tecnologias Avenida Campo Grande, 376, 1749-024 Lisboa, Portugal}

\date{\today}

\begin{abstract}
In this work we derive a matrix formulation of a noise-disturbance uncertainty relation, which is akin to the Robertson-Schr\"odinger uncertainty principle. Our inequality is stronger than Ozawa's uncertainty principle and takes noise-disturbance correlations into account. Moreover, we show that, for certain types of measurement interactions, it is covariant with respect to linear symplectic transformations of the noise and disturbance operators.
\end{abstract}

\maketitle

1. {\it Introduction} - Although, Heisenberg's uncertainty principle is one of the hallmarks of quantum mechanics, there has been some discussion about its formulation. Robertson's formulation of the uncertainty principle,
\be\label{eq1}
\sigma (A, \psi) \sigma (B, \psi) \ge {{| \langle \psi | \left[A, B \right] | \psi\rangle |}\over2}~,
\ee
expresses an intrinsic uncertainty of the states in terms of the standard deviation of some pair of noncommuting observables $A$ and $B$ in a state $\psi$. This kind of formulation describes a limitation on the preparation of the state, but has no direct relevance for the accuracy of the measurement of an observable $A$ with an apparatus and the disturbance caused by it on observable $B$. We shall refer to these formulations as kinematical uncertainty principles. 

Another kinematical inequality is the Robertson-Schr\"odinger uncertainty principle (RSUP). It can be stated in terms of the positivity of the matrix
\be
{\bf \Sigma} + {i \hbar\over2} {\bf J} \ge 0~,
\label{eq3}
\ee
where ${\bf \Sigma}$ is the covariance matrix of the state
\be
{\bf \Sigma} = \left(
\begin{array}{c c}
\sigma (X, \psi)^2 & \sigma (X,P, \psi)\\
\sigma (P, X, \psi) & \sigma (P, \psi)^2
\end{array}
\right).
\label{eq4}
\ee
Here $X$ is the particle position and $P$ the momentum, $\sigma (X,P, \psi)=\sigma (P,X, \psi) = \langle \psi | ~ \left\{ \Delta X, \Delta P \right\}~ | \psi\rangle $ are the covariance elements for position-momentum correlations, where $\left\{ \cdot, \cdot \right\}$ is the anti-commutator and $\Delta X = X- < \psi | X | \psi>$, etc. Also
\be
{\bf J} = \left(
\begin{array}{c c}
0& 1\\
- 1  & 0
\end{array}
\right)
\label{eq5}
\ee
is the standard symplectic matrix. This formulation has several advantages over the one stated in Eq. (\ref{eq1}). On the one hand, it is stronger than inequality (\ref{eq1}), in the sense that in fact it implies Eq. (\ref{eq1}). However, the converse is not true. On the other hand, it accounts for the position-momentum correlations, which are relevant in several contexts, e.g. states with strong position-momentum correlations may lead to greater transparency of the Coulomb barrier during the interaction of charged particles. This is quite relevant, for instance, in the astrophysics of stars and in controlled nuclear fusion, where the action of the Coulomb barrier leads to a very low tunneling probability for low-energy particles \cite{Vysotskii}. For Gaussian states (which include coherent, squeezed and thermal states), the RSUP constitutes the necessary and sufficient condition of quantumness. Moreover, after a certain reflection transformation \cite{Simon1}, it also establishes unequivocally the separable or entangled nature of Gaussian states. Experimentally, coherent and squeezed states play an important role in quantum optics \cite{Simon2}, quantum computation of continuous variables \cite{Giedke} and investigations of the quantum-classical transition \cite{Littlejohn}. Finally, the RSUP is invariant under linear symplectic transformations, a property which is not shared by inequality (\ref{eq1}). This is important in the context of semi-classical analysis \cite{Littlejohn} and in the search for directions of minimal uncertainty \cite{Narcowich}.

An experimental violation of the kinematical uncertainty principles, Eqs. (\ref{eq1}) and (\ref{eq3}) can only be attributed to either the failure of the Hilbert space formalism to correctly describe quantum systems - something which would have profound implications on the theoretical edifice of quantum mechanics -, or alternatively, to some modification of the position-momentum commutation relations. The latter possibility has been explored recently in Ref. \cite{Bastos1}. To the best of our knowledge, no such experimental violation has ever been recorded.

In addition to the previous kinematical inequalities, there are other uncertainty principles (dynamical uncertainty principles) which try to account for the ``unavoidable and uncontrollable disturbance" caused on observable $B$ by a measurement of observable $A$. In his famous $\gamma$-ray thought experiment \cite{Heisenberg}, Heisenberg argued that the product of the noise in a position measurement and the momentum disturbance caused by that measurement should be no less that ${\hbar/2}$. More generally, if $\epsilon (A, \psi) $ denotes the noise of the $A$ measurement and $\eta (B, \psi)$ the disturbance on $B$ caused by that measurement, when the system is in state $\psi$, then the Heisenberg noise-disturbance relation reads
\be
\epsilon (A, \psi) \eta (B, \psi) \ge {{|\langle \psi |  \left[A, B \right]  | \psi\rangle|}\over2}~.
\label{eq6}
\ee

There have been various accounts of theoretical \cite{Ozawa1,Ozawa2,Ozawa3} and experimental \cite{Rozema, Hasegawa} violations of inequality (\ref{eq6}). This has prompted the search for an universally valid formulation of an uncertainty principle accounting for the noise and disturbance of the measurement interaction. Recently, M. Ozawa \cite{Ozawa4} considered a composite system of the object and the measuring device (the probe), initially prepared in a product state $\Psi = \psi \otimes \xi$, where $\psi$ and $\xi$ describe the object and the probe, respectively. Working in the Heisenberg picture, he introduced the noise operator $N(A)$ associated with observable $A$ and the disturbance operator $D(B)$. They are self-adjoint operators, defined by
\be
N(A) = M^{out} - A^{in}\hspace{0,2cm}, \hspace{0,2cm} D(B) = B^{out} - B^{in}~.
\label{eq7}
\ee
Here $A^{in}=A \otimes I, B^{in}=B \otimes I$ are observables $A, B$ prior to the measurement interaction, $B^{out} =U^{\dagger} (B \otimes I) U$ is the observable $B$ immediately after the measurement and $M$ is the probe observable. $U$ is the unitary time evolution operator during the measuring interaction. Clearly, $M^{in} = I \otimes M$ and $M^{out} = U^{\dagger} (I \otimes M) U$. For more details on the measurement interaction see Ref. \cite{Ozawa4}. The noise $\epsilon (A, \psi)$ and disturbance $\eta (B; \psi)$ are defined to be the mean-square deviations of the noise and disturbance operators, respectively:
\be
\epsilon (A, \psi)^2 = \langle\Psi | \Delta N(A)^2 | \Psi\rangle\hspace{0,2cm}, \hspace{0,2 cm} \eta (B, \psi)^2 =  \langle \Psi | \Delta D(B)^2 | \Psi\rangle~.
\label{eq8}
\ee
Since $M$ and $B$ are observables in different systems, they commute : $\left[M^{out}, B^{out} \right]=0$. Using this fact, Eq.(\ref{eq8}), the triangle and the Cauchy-Schwartz inequalities, one can then prove Ozawa's uncertainty principle (OUP):
%\be
%\begin{array}{c}
%\epsilon (A, \psi) \eta (B, \psi)  +  {{\left| \langle \left[N(A),B^{in} \right]\rangle + \langle\left[A^{in} , D(B) \right]\rangle\right|}\over2} \ge \\
%\ge {{| \langle\psi| ~\left[A, B \right] ~| \psi\rangle}\over2}.
%\end{array}
%\label{eq9}
%\ee
%As in Ref. \cite{Ozawa4}, we use the abbreviated notation $\langleC\rangle$ to denote $\langle\Psi | C | \Psi\rangle$.
%If $\left|<\left[N(A),B^{in} \right]> + <\left[A^{in},D(B) \right]>\right|=0$, then the Heisenberg noise-disturbance uncertainty relation (\ref{eq6}) holds. A particular case corresponds to $\left[N(A),B^{in} \right] = \left[A^{in} , D(B) \right]=0$. Ozawa defined such a measuring interaction to be of {\it independent intervention} for the pair $(A,B)$.
%From the triangle and the Cauchy-Schwartz inequalities one has:
%\be
%\begin{array}{c}
%\left| < \left[N(A),B^{in} \right]> + < \left[A^{in} , D(B) \right]>\right|  \le  \\
%\le 2 \epsilon (A, \psi) \sigma (B, \psi) + 2 \sigma (A, \psi) \eta (B, \psi).
 %\end{array}
%\label{eq10}
%\ee
%Upon substitution of (\ref{eq10}) in (\ref{eq9}) one obtains:
\be
\!\!\!\!\!\epsilon (A, \psi) \eta (B, \psi)\! +\!\epsilon (A, \psi) \sigma (B, \psi)\!+\!\sigma (A, \psi) \eta (B, \psi)\!\! \ge\!\! {{| \langle\psi|\!\! \left[A, B \right]\!\! | \psi\rangle}\over2}~.
\label{eq11}
\ee
Just as the kinematical uncertainty principle, Eq. (\ref{eq1}), does not account for the position-momentum correlations, neither does the OUP, Eq. (\ref{eq11}), account for the noise-disturbance correlations.

The purpose of this letter is then to derive a matrix formulation of the OUP, which encompasses the noise-disturbance correlations. It is related to the OUP in very much the same way as the RSUP relates to Eq. (\ref{eq1}) as it is more general than the latter and admits nicer symmetry properties.
%The paper is organized as follows. In the next section, we derive our matrix formulation of the noise-disturbance uncertainty principle. In section 3, we prove that the OUP is a consequence of our matrix formulation and we give a counter-example that shows that the converse is not true. In section 4, we discuss the covariance properties of the new uncertainty principle under a certain class of unitary transformations.

2. {\it Matrix formulation of the Ozawa uncertainty principle} - In the sequel, Latin indices $i,j$ run in the set $\left\{1, \cdots, n\right\}$ and Greek indices $\alpha, \beta$ in the set $\left\{1, \cdots, 2n\right\}$. For the sake of generality, we shall also consider a multidimensional system. Using Ozawa's notation, let $A_i^{in}$ and $B_j^{in}$, $i,j=1, \cdots, n$ denote some set of self-adjoint operators such that
\be
\left[A_i^{in},A_j^{in} \right] = \left[B_i^{in},B_j^{in} \right] =0\hspace{0,2cm}, \hspace{0,2cm} \left[A_i^{in}, B_j^{in} \right] = i C_{ij}~,
\label{eq12}
\ee
for $i,j=1, \cdots, n$, and where $\left\{C_{ij} \right\}$ are some self-adjoint operators. If $A$ and $B$ are the particle's position and momentum, we simply have $C_{ij} = \hbar \delta_{ij} $. We may write these collectively as $Z^{in} = \left(A_1^{in}, \cdots, A_n^{in},B_1^{in}, \cdots, B_n^{in} \right)$ satisfying the commutation relations
\be
\left[Z_{\alpha}^{in},Z_{\beta}^{in} \right] = i G_{\alpha \beta}\hspace{0,2cm}, \hspace{0,2 cm} \alpha, \beta =1, \cdots, 2n~.
\label{eq14}
\ee
${\bf G} = \left\{G_{\alpha \beta} \right\}$ is the self-adjoint operator-valued skew-symmetric matrix
\be
{\bf G} = \left(
\begin{array}{c c}
{\bf 0} & {\bf C}\\
- {\bf C} & {\bf 0}
\end{array}
\right),
\label{eq15}
\ee
with ${\bf C} = \left\{C_{ij} \right\}$. Again, if $A$ and $B$ are the position and momentum operators, then we simply have $2n \times 2n$ standard symplectic matrix $\hbar {\bf J}$.

Let us define the noise and disturbance operators as
\ba
N  &=& N (A) = \left(N_1, \cdots, N_n \right)~,\\
D  &=& D (B) = \left(D_1, \cdots, D_n \right)~.
\label{eq16}
\ea
We can write these collectively as
\be
K = \left(N_1, \cdots, N_n, D_1, \cdots, D_n \right)~.
\label{eq18}
\ee
Then, we denote the output of the (commuting) probe observables $M^{out}$ and the output of $B$ as
\ba
M^{out} &=& \left(M^{out}_1, \cdots, M^{out}_n \right)~,\\
B^{out} &=& \left( B_1^{out}, \cdots, B_n^{out} \right)~.
\label{eq19}
\ea
If we write $Z^{out} = \left(M^{out}_1, \cdots, M^{out}_n,B_1^{out}, \cdots, B_n^{out} \right)$ then we have as before:
\be
\left[Z_{\alpha}^{out}, Z_{\beta}^{out} \right]=0\hspace{0,2 cm}, \hspace{0,2 cm} \alpha , \beta =1, \cdots, 2n~,
\label{eq22}
\ee
and $Z^{out}= Z^{in} + K$.

Let $\left\{\lambda_{\alpha} \right\}_{1 \le \alpha \le 2n}$ denote an arbitrary set of complex numbers. Thus, we have:
\ba
 &&\!\!\!\!\!\! \sum_{\alpha, \beta =1}^{2n}\!\!\!\! \overline{\lambda}_{\alpha} \lambda_{\beta} \langle \left[\Delta Z_{\alpha}^{out},  \Delta Z_{\beta}^{out} \right]\rangle\nonumber\\
%&&=\sum_{\alpha, \beta =1}^{2n} \overline{\lambda}_{\alpha} \lambda_{\beta} \langle\left[\Delta Z_{\alpha}^{in} + \Delta K_{\alpha} ,\Delta Z_{\beta}^{in} + \Delta K_{\beta} \right]\rangle \nonumber\\
&&\!\!\!\!\!\!=\!\!\!\!\!\!\sum_{\alpha, \beta =1}^{2n}\!\!\!\! \overline{\lambda}_{\alpha} \lambda_{\beta}\!\! \left(i \langle G_{\alpha \beta}\rangle\!\! +\!\! \langle \left[ Z_{\alpha}^{in},K_{\beta} \right]\!\!+\!\!  \left[ K_{\alpha} ,Z_{\beta}^{in} \right]\rangle\!\!+\!\!\langle \left[\Delta K_{\alpha} ,\Delta K_{\beta} \right]\rangle\!\right)\nonumber\\
&&\!\!\!\!\!\!=0~.
\label{eq24}
\ea
Now notice that writing $\mathcal{K} = \sum_{\alpha}\lambda_{\alpha} \Delta K_{\alpha}$, we have
\be
\!\!\!\!\sum_{\alpha, \beta =1}^{2n} \overline{\lambda}_{\alpha} \lambda_{\beta}  \langle \left[\Delta K_{\alpha} ,\Delta K_{\beta} \right]\rangle = \langle \mathcal{K}^{\dagger} \mathcal{K} \rangle- \langle\mathcal{K} \mathcal{K}^{\dagger} \rangle \le\!\!  \langle \mathcal{K}^{\dagger} \mathcal{K}\rangle~,
\label{eq25}
\ee
and so
\ba
&&\!\!\!\!\!\!\!\!\!\sum_{\alpha, \beta =1}^{2n}\!\!\!\! \overline{\lambda}_{\alpha} \lambda_{\beta}  \langle\left[\Delta K_{\alpha} ,\Delta K_{\beta} \right]\rangle \le \!\!\! \sum_{\alpha, \beta =1}^{2n} \overline{\lambda}_{\alpha} \lambda_{\beta} \langle \Delta K_{\alpha} \Delta K_{\beta}\rangle \nonumber\\
&&\!\!\!\!\!\!\!\!\!=\sum_{\alpha, \beta =1}^{2n}\!\!\!\! \overline{\lambda}_{\alpha} \lambda_{\beta}\!\!\left(\!\! \langle\left\{\Delta K_{\alpha} ,\Delta K_{\beta} \right\}\rangle + {1\over2} \langle \left[\Delta K_{\alpha} ,\Delta K_{\beta} \right]\rangle\!\! \right)~.
\label{eq26}
\ea
Thus,
\be
\!\!\!\!\!\!\sum_{\alpha, \beta =1}^{2n}\!\!\!\! \overline{\lambda}_{\alpha} \lambda_{\beta}  \langle \left[\Delta K_{\alpha} ,\Delta K_{\beta} \right]\rangle \le 2\!\! \sum_{\alpha, \beta =1}^{2n}\!\!\!\! \overline{\lambda}_{\alpha} \lambda_{\beta}\langle \left\{\Delta K_{\alpha} ,\Delta K_{\beta} \right\}\rangle~.
\label{eq27}
\ee
Upon substitution of Eq. (\ref{eq27}) into Eq. (\ref{eq24}) we obtain
\ba
&&\!\!\!\!\!\!\!\!\sum_{\alpha, \beta =1}^{2n}\!\!\!\! \overline{\lambda}_{\alpha} \lambda_{\beta}\!\!\left(i \langle G_{\alpha \beta}\rangle\!\!+\!\! \langle\left[ Z_{\alpha}^{in},K_{\beta} \right]\!\! + \!\! \left[K_{\alpha} ,Z_{\beta}^{in} \right]\rangle\!\! +\!\!  2 \langle \left\{\Delta K_{\alpha} ,\Delta K_{\beta} \right\}\rangle\right)\nonumber\\
&&\!\!\!\!\!\!\!\!\ge0~.
\label{eq28}
\ea
If we define the $2n \times 2n$ real symmetric positive-definite matrix ${\bf K}_{\alpha \beta} =   \langle \left\{\Delta K_{\alpha} ,\Delta K_{\beta} \right\}\rangle$, the $2n \times 2n$ real skew-symmetric matrices ${\bf \mathcal{G}} = \langle{\bf G}\rangle$ and
\be
{\bf \Gamma}_{\alpha \beta} = {1\over i}  \langle \left[ Z_{\alpha}^{in},K_{\beta} \right] +  \left[K_{\alpha} ,Z_{\beta}^{in} \right]\rangle,
\label{eq30}
\ee
then we can rewrite Eq. (\ref{eq28}) in the matrix form:
\be
{\bf K} + {i\over2} \left( {\bf \Gamma} + {\bf \mathcal{G}} \right) \ge 0,
\label{eq31}
\ee
which is our matrix formulation of Ozawa's uncertainty principle. In Ozawa's terminology, if the measuring interaction is of {\it independent intervention} \cite{Ozawa4}, i.e. if
\be
{\bf \Gamma} =0,
\label{eq32}
\ee
then we obtain
\be
{\bf K} + {i\over2} {\bf \mathcal{G}}  \ge 0,
\label{eq33}
\ee
which is the matrix generalization of the Heisenberg noise-disturbance relation, Eq. (\ref{eq6}), based on the $\gamma$-ray thought experiment.

3. {\it On the connection with the Ozawa uncertainty principle} - Here, we argue that our formulation of the uncertainty principle, Eq. (\ref{eq31}), is in fact stronger than Ozawa's uncertainty principle. Indeed, let us consider for simplicity $n=1$. We then have
\be
{\bf K} = \left(
\begin{array}{c c}
\langle\left\{\Delta N, \Delta N \right\}\rangle & \langle\left\{\Delta N, \Delta D \right\}\rangle \\
\\
\langle\left\{\Delta D, \Delta N \right\}\rangle & \langle\left\{\Delta D, \Delta D \right\}\rangle
\end{array}
\right),
\label{eq34}
\ee
while
\be
{\bf \Gamma} = \frac{1}{i} \left(
\begin{array}{c c}
0 & \langle\left[ A^{in},  D \right] +\left[N, B^{in}   \right] \rangle \\
\\
\langle\left[  D , A^{in}\right] +\left[ B^{in}, N   \right]  \rangle & 0
\end{array}
\right).
\label{eq35}
\ee

If Eq. (\ref{eq31}) holds, then the matrix ${\bf K} + \frac{i}{2} \left( {\bf \Gamma} + {\bf \mathcal{G}} \right)$ must have non-negative determinant, and we obtain
\ba
&&\!\!\!\!\!\langle\left\{\Delta N, \Delta N \right\}\rangle\langle\left\{\Delta D, \Delta D \right\}\rangle\nonumber\\
&&\!\!\!\!\!\ge\langle\left\{\Delta N, \Delta D \right\}\rangle^2\!\! +\!\!{1\over4} \left| \langle\left[ A^{in},  D \right] \!\!+\!\!\left[N, B^{in}   \right]  \rangle\!\! +\!\! \langle \left[A^{in}, B^{in} \right]\rangle \right|^2\nonumber\\
&&\!\!\!\!\!\ge {1\over4} \left| \langle\left[ A^{in},  D \right] +\left[N, B^{in}   \right]  \rangle + \langle \left[A^{in}, B^{in} \right]\rangle \right|^2~.
 \label{eq36}
\ea
Taking the square root and writing $\epsilon (A) = \langle\left\{\Delta N, \Delta N \right\}\rangle^{1/2}$ and $\eta (B) = \langle\left\{\Delta D, \Delta D \right\}\rangle^{1/2}$, and using the inequality $|a-b| \ge \left|~ |a|- |b|  ~ \right|$, we finally obtain
%\be\begin{array}{c}<\left\{\Delta N, \Delta N \right\}>^{1/2} <\left\{\Delta D, \Delta D \right\}>^{1/2}  ~ \ge \\\\\ge ~  \frac{1}{2} \left| <\left[ A^{in},  D \right] +\left[N, B^{in}   \right]  > - < \left[B^{in}, A^{in} \right]> \right|\end{array}\label{eq37}\ee
\be
\epsilon (A) \eta (B) \ge {1\over2} \left| ~| \langle\left[ A^{in},  D \right] +\left[N, B^{in}   \right]  \rangle| - |  \langle \left[A^{in}, B^{in} \right]\rangle  | ~\right|~.
\label{eq38}
\ee
In particular
\be
\epsilon (A) \eta (B) \ge {1\over2} |  \langle\left[A^{in}, B^{in} \right]\rangle  | - {1\over2} \left|  \langle\left[ A^{in},  D \right] +\left[N, B^{in}   \right]  \rangle\right|~,
\label{eq39}
\ee
which is Ozawa's uncertainty principle before using the triangular identity and the Cauchy-Schwartz relation. The obvious question is now whether our uncertainty, Eq. (\ref{eq31}), is equivalent to Ozawa's or whether it is in fact more restrictive. The latter is true and in the following we show it using a type of measuring interaction known as a {\it backaction evading quadrature amplifier} \cite{Ozawa4,Yurke}. In this case, the system is described by a set of quadrature operators $(X_a,Y_a)$ and the probe by the operators $(X_b,Y_b)$ with
\be
\left[X_a,Y_a \right]= \left[X_b,Y_b \right] = {i\over2}~.
\label{eq40}
\ee
Then we have the measuring interaction
\be
\left\{
\begin{array}{l}
X_a^{out}= X_a^{in}~,\\
X_b^{out} = X_b^{in} +  G X_a^{in}~,\\
Y_a^{out} = Y_a^{in} - G Y_b^{in}~,\\
Y_b^{out} = Y_b^{in}~,
\end{array}
\right.
\label{eq41}
\ee
where $G$ is the gain. The probe observable is then set to $M= (1/G) X_b$, and thus
\be
M^{out} = X_a^{in} + {1\over G} X_b^{in}~.
\label{eq42}
\ee
Moreover
\be
\begin{array}{l}
N (X_a)= \frac{1}{G} X_b^{in}~,\\
D (X_a) =0~,\\
D (Y_a) = - G Y_b^{in}~.
\end{array}
\label{eq43}
\ee
Following our previous notation, we set $A^{in} = X_a^{in}$, $B^{in} = Y_a^{in}$. Then, $Z^{in} = \left(X_a^{in}, Y_a^{in} \right)$
and given that
\be
N =N (X_a) = {1\over G} X_b^{in}\hspace{0,2cm}, \hspace{0,2 cm} D =D (Y_a) = -G Y_b^{in}~,
\label{eq44}
\ee
one can write
\be\label{eq45}
K = \left({1\over G} X_b^{in}, -G Y_b^{in} \right)~.
\ee
We conclude that ${\bf \Gamma } =0$ and that the measuring interaction is of independent intervention for the pair $(X_a,Y_a)$.
Also from (\ref{eq40}):
\be
{\bf \mathcal{G}} = \frac{1}{2} {\bf J}.
\label{eq46}
\ee

Now let us consider the state of the probe $\xi$ to have a covariance matrix
\be
{\bf \Sigma^b} = \left(
\begin{array}{c c}
{\bf \Sigma^b}_{11} & {\bf \Sigma^b}_{12} \\
& \\
{\bf \Sigma^b}_{12} & {\bf \Sigma^b}_{22}
\end{array}
\right) =
\left(
\begin{array}{c c}
{1\over4} & {1\over2} \\
& \\
{1\over2} & {1\over4}
\end{array}
\right).
\label{eq47}
\ee
Next, notice that from Eq. (\ref{eq45}):
\be
\langle \left\{N,N \right\}\rangle = {1\over G^2} \langle \left\{X_b^{in},X_b^{in} \right\}\rangle = {1\over G^2} {\bf \Sigma^b}_{11} ={1\over 4 G^2}~.
 \label{eq48}
\ee
In a similar fashion, we obtain
\ba
\langle \left\{N,D \right\}\rangle =-\langle \left\{X_b^{in},Y_b^{in} \right\}\rangle = - {1\over2}~,\\
\langle\left\{D,D \right\}\rangle = G^2 \langle\left\{Y_b^{in},Y_b^{in} \right\}\rangle = {G^2\over4}~.
 \label{eq50}
\ea
Thus,
\be
{\bf \Sigma} = \left(
\begin{array}{c c}
{1\over4 G^2}  & - {1\over2}\\
& \\
- {1\over2} &  {G^2\over4}
\end{array}
\right)
\label{eq51}
\ee
We have then from Eqs. (\ref{eq46}) and (\ref{eq51})
\be
{\bf K} + {i\over2} \left({\bf \Gamma} + {\bf \mathcal{G}} \right)={\bf \Sigma} + {i\over4} {\bf J} = \left(
\begin{array}{c c}
{1\over4 G^2}  & - {1\over2} + {i\over4}\\
& \\
- {1\over2} -{i\over4} &  {G^2\over4}
\end{array}
\right)~.
\label{eq52}
\ee
Since $\det ({\bf K} + {i\over4} {\bf J}) = - {1\over4}$, we conclude that it is not a positive matrix and that our uncertainty principle Eq. (\ref{eq31}) is violated. However, since $\epsilon (A) = \langle \left\{ N, N \right\} \rangle^{1/2}= {1\over2G}$ and $\eta (B) = \langle\left\{ D,D \right\} \rangle^{1/2}= {G\over2}$, we get that $\epsilon (A)\eta (B) = {1\over4}$, which exactly saturates the OUP. This proves that our inequality (\ref{eq31}) is indeed stronger than OUP, Eq. (\ref{eq11}).

4. {\it Comments on the invariance properties of the matrix formulation} - The matrix formulation of the noise-disturbance uncertainty principle Eq. (\ref{eq31}) is universally applicable. Given the myriad of measurement interactions and apparatuses \cite{Wheeler,Ozawa5}, it is virtually impossible to establish all transformations which leave Eq. (\ref{eq31}) unchanged. There are nonetheless instances, where the inequality (\ref{eq31}) is preserved under a certain type of transformation, while Eq. (\ref{eq11}) is not.

Suppose that $\left[ K_{\alpha} , K_{\beta} \right] = i \gamma J_{\alpha \beta}$, where $\gamma \ne 0$ is some real constant and ${\bf J}$ is the standard symplectic matrix. Moreover, let the measurement interaction be of independent intervention (${\bf \Gamma} =0$). Then Eq. (\ref{eq31}) becomes ${\bf K} + {i \gamma\over2} {\bf J} \ge 0$. Notice that this looks formally like the RSUP, Eq. (\ref{eq3}). Suppose that the system undergoes a linear symplectic transformation $K_{\alpha} \mapsto K_{\alpha}^{\prime} = \sum_{1 \le \beta \le 2n} S_{\alpha \beta} K_{\beta}$, where ${\bf S} \in Sp(2n; \mathbb{R})$. Then the noise-disturbance covariance matrix transforms by similarity ${\bf K} \mapsto {\bf K}^{\prime} = {\bf S}{\bf K} {\bf S}^T$. But given that ${\bf S}^{-1} {\bf J} ({\bf S}^{-1})^T = {\bf J}$, we conclude that the matrix uncertainty principle remains unchanged: ${\bf K}^{\prime} + {i \gamma\over2} {\bf J} \ge 0$.

But this may not happen for the Ozawa uncertainty relation. Indeed, let us consider again our previous example of the backaction evading quadrature amplifier. Remember that, for an interaction of independent intervention, the Ozawa inequality becomes simply the Heisenberg inequality
\begin{equation}
\epsilon \eta \ge {1\over4}~.
 \label{eq53}
 \ee
Suppose that we have now a noise-disturbance covariance matrix ${\bf K} = diag (\epsilon^2,\eta^2)$ and that the probe (and possibly the object) is subjected to a symplectic transformation such that:
\be
K \mapsto K^{\prime} = {\sqrt{2}\over2} \left(
\begin{array}{c c}
1 & 1\\
-1 & 1
\end{array}
\right) K.
\label{eq54}
\ee
That is: the noise-disturbance vector $K$ is rotated through an angle of $\frac{\pi}{4}$. Such a transformation is easily implemented by a certain unitary transformation $U(S)$ generated by an appropriate hermitian operator, quadratic in the variables $X_b, Y_b$ of the probe.

Then the Ozawa uncertainty inequality is modified to
\be
(\epsilon^{\prime} )^2 + (\eta^{\prime} )^2  \ge \sqrt{ 1 + 4 \langle \left\{\Delta N^{\prime} ,\Delta D^{\prime} \right\} \rangle^2},
\label{eq55}
\ee
which is manifestly different from Eq. (\ref{eq53}). We also remark that the noise-disturbance correlations naturally appear under such transformations.

5. {\it Conclusions} - In this work we presented a universal matrix formulation of the uncertainty principle which is more stringent than the noise-disturbance relation of Ozawa. Indeed, we have proved that our formulation implies Ozawa's, and showed that is possible to saturate the Ozawa uncertainty principle, while violating our universal form. Our inequality is also more general in the sense that, unlike Ozawa's relation, it is also accounts for the noise-disturbance correlations.

We recall that recent experimental work performed by Rozema et al. \cite{Rozema}, using polarized entangled photons, and by Sulyok et al. \cite{Hasegawa}, using neutron-spin measurements, proved the validity of Ozawa's relation using weak measurements. It would certainly be an interesting prospect to investigate the experimental validity of our relation with a similar experimental setup.

\begin{acknowledgements}
The work of CB is supported by the FCT (Portugal) grant SFRH/BPD/62861/2009. The AEB work is supported by the FAPESP (Brazil) grant 2012/03561-0. The work of OB is partially supported by the FCT project PTDC/FIS/111362/2009. NCD and JNP have been supported by the FCT grant PTDC/MAT/099880/2008.
\end{acknowledgements}


\begin{thebibliography}{99}



\bibitem{Vysotskii} V.I. Vysotskii, M.V. Vysotskyy, S.V. Adamenko, JETP 114 (2012) 243; J. Surf. Inv., X-Ray, Journal of Surface Investigation. Xray, Synchrotron and Neutron Techniques. 6 (2012) 369; V.I. Vysotskii, S.V. Adamenko, Tech. Phys. 55 (2010) 613.

\bibitem{Simon1} R. Simon, Phys. Rev. Lett. 84 (2000) 2726.

\bibitem{Simon2} R. Simon, E.C.G. Sudarshan, N. Mukunda, Phys. Rev. A 36 (1987) 3868.

\bibitem{Giedke} G. Giedke: Quantum information and continuous variable systems. PhD Thesis (Innsbruck, 2001).

\bibitem{Littlejohn} R.G. Littlejohn, Phys. Rep. 138 (1986) 193.

\bibitem{Narcowich} F.J. Narcowich, J. Math. Phys. 31 (1990) 354.

\bibitem{Bastos1} C. Bastos, O. Bertolami, N.C. Dias, J.N. Prata, Phys. Rev. D 86 (2012) 105030.

\bibitem{Heisenberg} W. Heisenberg, Z. Phys. 43 (1927) 172.

\bibitem{Ozawa1} M. Ozawa, Phys. Rev. Lett. 60 (1988) 385.

\bibitem{Ozawa2} M. Ozawa, in {\it Squeezed and Nonclassical Light}, edited by P. Tombesi and E.R. Pike (Plenum, New York, 1989) pp. 263-268.

\bibitem{Ozawa3} M. Ozawa, Phys. Lett. A 299 (2002) 1.

\bibitem{Rozema} L. A. Rozema, A. Darabi, D. H. Mahler, A. Hayat, Y. Soudagar, and A. M. Steinberg, Phys. Rev. Lett. 109, 100404 (2012).

\bibitem{Hasegawa} G. Sulyok, S. Sponar, J. Erhart, G. Badurek, M. Ozawa and Y. Hasegawa, Phys. Rev. A 88, 022110 (2013).

\bibitem{Ozawa4} M. Ozawa, Phys. Rev. A 67 (2003) 042105.

\bibitem{Yurke} B. Yurke, J. Opt. Soc. Am. B 2 (1985) 732.

\bibitem{Wheeler} {\it Quantum Theory and Measurement}, edited by J.A. Wheeler and W.H. Zurek (Princeton University Press, Princeton, 1983).

\bibitem{Ozawa5} M. Ozawa, Ann. Phys. 311 (2004) 350.



\end{thebibliography}
\end{document}